\documentclass[conference]{IEEEtran}
\IEEEoverridecommandlockouts
\usepackage{cite}
\usepackage{amsmath,amssymb,amsfonts}
\usepackage{algorithmic}
\usepackage{graphicx}
\usepackage{textcomp}
\usepackage{xcolor}
\usepackage{tabularx}
\usepackage{makecell}
\usepackage{subcaption}
\usepackage{adjustbox}
\usepackage{xcolor}
\usepackage{hyperref}
\usepackage{soul}

\def\BibTeX{{\rm B\kern-.05em{\sc i\kern-.025em b}\kern-.08em
    T\kern-.1667em\lower.7ex\hbox{E}\kern-.125emX}}
\begin{document}

\title{
Digital Twin-Empowered Routing Management for Reliable Multi-Hop Millimeter Wave V2X

}

\author{
    \IEEEauthorblockN{Supat Roongpraiwan, Zongdian Li, 
    Tao Yu, and
    Kei Sakaguchi}
    \IEEEauthorblockA{Department of Electrical and Electronic Engineering, Tokyo Institute of Technology, Tokyo, Japan.}  
    \IEEEauthorblockA{Emails: \{supat, lizd, yutao, sakaguchi\}@mobile.ee.titech.ac.jp} 

    }

\maketitle

\begin{abstract}
Digital twin (DT) technology can replicate physical entities in cyberspace.
A mobility DT digitalizes connected and autonomous vehicles (CAVs) and their surrounding traffic environment, allowing to monitor the maneuvering and distribution of CAVs in real-time, which is crucial for managing vehicle-to-everything (V2X) connectivity, especially when millimeter wave (mmWave) is adopted. MmWave V2X relies on dynamic multi-hop communications to ensure high reliability. Therefore, in this paper, the challenges of mmWave V2X are presented to motivate the utilization of DT, and then we introduce the system model for DT-based multi-hop routing management, incorporating two different routing algorithms: with and without future trajectory prediction. For proof of concept, we implement the proposed DT system using Unity-based AWSIM and evaluate the proposed algorithms via simulations. The results show that, compared to the conventional routing algorithm in vehicular ad hoc networks (VANETs), the DT-based algorithms significantly improve the reliability of mmWave V2X, and such improvements can be seen in both fully connected and mixed traffic scenarios.
\end{abstract}

\section{Introduction}
Vehicle-to-everything (V2X) communication is a critical component for connected and autonomous vehicles (CAVs) and intelligent transportation systems (ITS). By networking vehicles, pedestrians, and roadside units (RSUs) via vehicle-to-vehicle (V2V), vehicle-to-pedestrian (V2P), and vehicle-to-infrastructure (V2I) communications, respectively, a wide range of V2X applications for road safety, traffic efficiency, and entertainment have been empowered, such as forward collision warning (FCW), green light optimal speed advisory (GLOSA), and vehicle near-field payment (VNFP) \cite{cite:v2x-services}. Amid them, the safety-critical V2X applications are considered as the key enablers of level 4 - 5 autonomous vehicles \cite{cite:saej3016}.

Nowadays, millimeter wave (mmWave) is at the center of attention as it can significantly enhance V2X communication. The sufficient mmWave bandwidth within 30 GHz - 300 GHz unleashes the limitations of V2X communication to achieve multi-gigabit data rate and sub-millisecond latency \cite{cite:mmWave-V2X}. Such extraordinary performance will enable many advanced V2X applications, such as cooperative perception, which allows the exchange of extensive sensor data from light detection and rangings (LiDARs) and cameras in real time \cite{cite:cooperative-perception}.

However, one of the long-standing obstacles that hinder mmWave for V2X is reliability \cite{cite:mmwave-v2x-challenge}. As specified in the 3rd generation partnership project (3GPP) technical specification (TS) 22.185 \cite{cite:3gpp.22.185} and 22.186 \cite{cite:3gpp.22.186}, the reliability requirements for basic and advanced V2X applications range from 90\% - 99.999\%. Traditional V2X communications using lower frequencies around 5.9 GHz, such as the dedicated short-range communications (DSRC) and long-term evolution (LTE)-V2X, can satisfy such requirements via broadcasting-based routing schemes \cite{cite:v2x-broadcasting}. In contrast, mmWave V2X has short communication coverage due to severe propagation loss. The directivity and weak penetration capability of mmWave V2X make it prone to blockages. Hence, conventional broadcasting-based routing is no longer suitable for mmWave V2X to ensure high reliability. Instead, mmWave V2X needs efficient multi-hop management in dynamic traffic environments.

Thanks to the rapid development of internet of things (IoT) technology, digital twin (DT) emerges as a promising tool to monitor the real-time status and evolution of the physical world from cyberspace \cite{cite:DT-trend}. In \cite{cite:mobility-dt}, a mobility DT able to visualize dynamic traffic objects, including vehicles, cyclists, and pedestrians, with stationary environment models of roads, buildings, and trees, was established by using advanced ITS networks. Observing the mobility DT's potential in providing a global view of CAV networks with detailed environment information, this paper proposes to utilize such a DT for managing mmWave V2X in dynamic multi-hop communications. Our efforts focus on the following aspects:
\begin{itemize}
    \item We introduce the system architecture of DT-based multi-hop routing management for mmWave V2X. We define the mmWave V2X channel model and reliability metric, and describe the functional components in detail.
    \item We propose two different DT-based multi-hop routing algorithms: without and without vehicle trajectory prediction. The one without trajectory prediction indicates the ideal case that the DT is perfectly synchronized, while the other exploits trajectory prediction to compensate for the mis-synchronization due to network delay and fluctuation.
    \item We conduct a proof of concept to evaluate our proposed routing algorithms. Firstly, we utilize the open-sourced software AWSIM \cite{cite:awsim} to establish a mobility DT of the Nishishinjuku Area. Then, the mmWave V2X topology is extracted for routing calculation. Finally, we compare the reliability versus the conventional routing method in vehicular ad hoc networks (VANETs).
\end{itemize}


The remainder of this paper is organized as follows: Sect. \ref{section:related-works} presents some related works and reveals their limitations. Sect. \ref{section:DT-based-multi-hop} provides the details of the DT-based multi-hop routing management system for mmWave V2X.
Sect. \ref{section:evaluation} introduces the system implementation and evaluates our proposed routing algorithms via simulations.
Finally, Sect. \ref{section:conclusion} concludes the paper and discusses the potential future work.


\section{Related Works\label{section:related-works}}
\subsection{mmWave routing algorithms}
Beamforming is needed for mmWave to overcome the high propagation loss. However, directional beams lead to unstable connections in highly dynamic mobile networks. Due to this constraint, the mmWave is so far primarily used for backhaul networks, which have a stationary network topology. Classical ad-hoc routing algorithms, such as ad-hoc on-demand distance vector routing (AODV), dynamic source routing (DSR), and optimized link state routing (OLSR), still function well in the mmWave mesh backhaul networks, although some researchers investigated new routing algorithms to minimize energy consumption \cite{cite:rw-a-1}, reduce network latency \cite{cite:rw-a-2}, and enable fast link failure recovery \cite{cite:rw-a-3}. Nevertheless, these algorithms will suffer performance degradation in mmWave mobile networks due to the lack of mobility considerations. In \cite{cite:rw-a-4}, the researchers proposed a proactive route refinement scheme for mmWave, which can mitigate the impacts of mobility and human blockage, but their simulation only assumes an indoor environment. Therefore, it is questionable whether this scheme can perform well in a more dynamic outdoor environment.

Indeed, a few researchers have worked on the routing algorithms for mmWave-based vehicular networks and considered beam alignment issues \cite{cite:rw-a-5, cite:rw-a-6}. Their common characteristic is requiring accurate vehicle positions as side information. Although such information can be broadcast by vehicles within the VANET, the delay is hard to predict due to the carrier-sense multiple access with collision avoidance (CSMA/CA) mechanism. Therefore, these algorithms need to be improved with more efficient and reliable vehicle position/topology acquisition methods to adapt to the practical V2X environment.

\subsection{Software-defined vehicular networks (SDVNs)}
The decentralized network architecture is one of the major reasons that global information, like node status, is difficult to collect from VANETs. Such information is critical for multi-hop routing in terms of topology acquisition. In this respect, software-defined networking (SDN) has advantages because its centralized architecture design makes the network control plane functions, like routing management, more noticeable, more robust, and more responsive to data plane dynamics \cite{cite:rw-b-1}. Therefore, software-defined vehicular networks (SDVNs) have been proposed as a novel V2X network architecture \cite{cite:rw-b-2, cite:rw-b-3, cite:rw-b-4}. In fact, the discussions about SDVNs are not limited to theoretical and simulation levels. Some researchers have built their SDVN testbeds \cite{cite:rw-b-5, cite:rw-b-6}, and some even conducted field trials for SDVN-based multi-hop mmWave V2X \cite{cite:rw-b-7, cite:rw-b-8}.

Although SDVNs can efficiently generate/update global network topology, it can only collect information from connected entities, such as CAVs and roadside units (RSUs). Apparently, such information is not sufficient for mmWave multi-hop routing management because non-connected vehicles and physical objects like trees, buildings, etc., can influence V2X links. In other words, environmental information should be as detailed as possible in the topology. The afore-mentioned DT owns 3D models for all those elements, regardless of the connected or non-connected and the dynamic or static. Therefore, DT must be a critical component for mmWave V2X systems.

\section{DT-based Multi-Hop Routing Management\label{section:DT-based-multi-hop}}
\begin{figure}[t]
    \includegraphics[width=\linewidth]{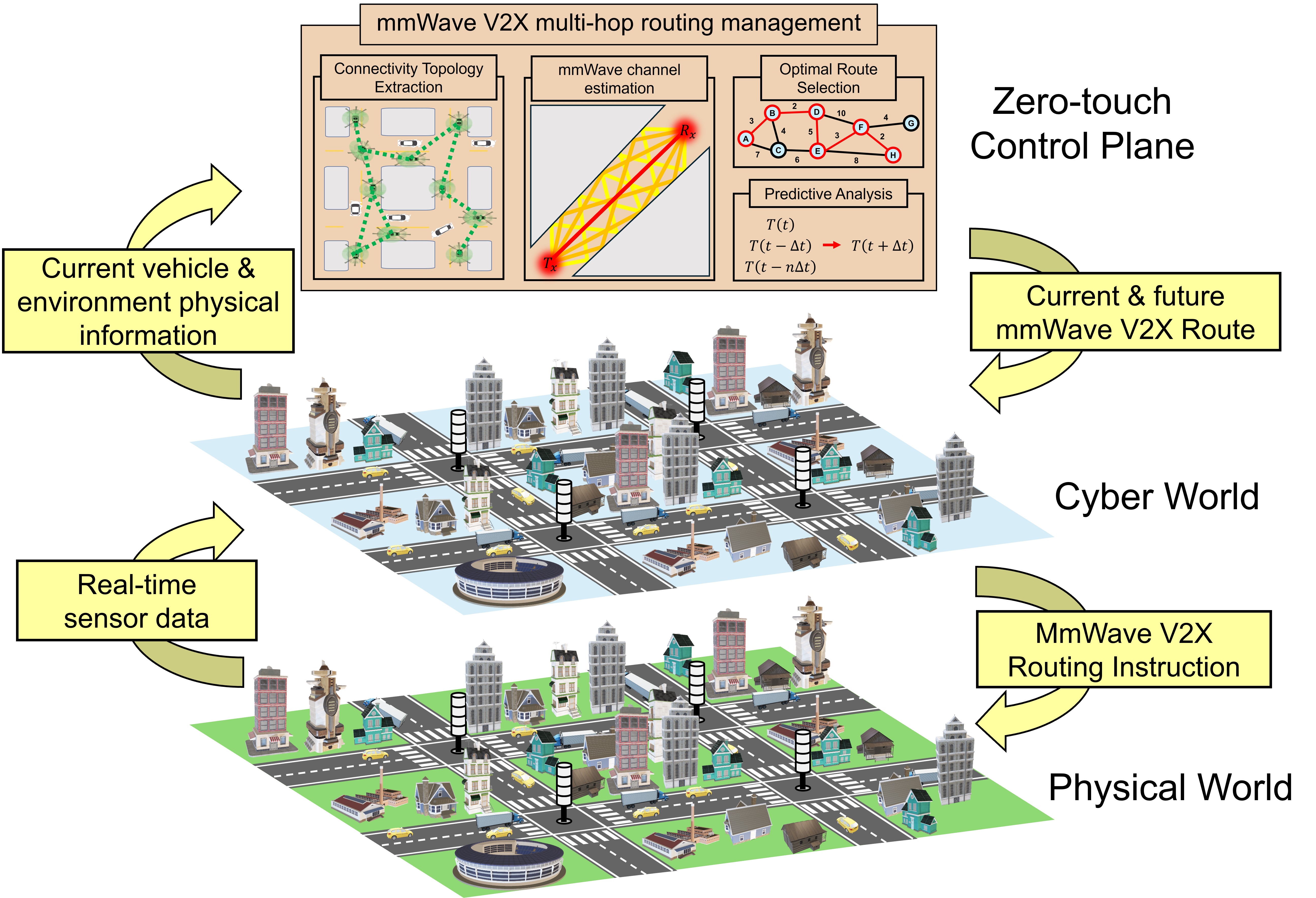}
    \centering
    \caption{System architecture.}
    \label{fig:sys-arch}
\end{figure}
\subsection{System architecture}
This section introduces a system architecture that leverages DT technology for managing reliable multi-hop mmWave V2X communication within dynamic environments. As illustrated in Fig. \ref{fig:sys-arch}, this system architecture provides a zero-touch control plane for managing mmWave multi-hop routing communication, utilizing real-time dynamic information from the DT.

The DT of the vehicle system comprises both the physical and cyber world. The cyber world model is initially constructed using location-specific static data, including high-definition 3D maps, 3D models, and vector maps containing detailed traffic and road infrastructure information. As real-time sensor data is gathered from RSUs and CAVs, the data is sent to DT server via V2X communication, enabling the cyber world to dynamically reflect the vehicle's real-time status, including its position and shape. This static and dynamic data integration ensures that the DT cyber world can accurately represent the vehicle's behavior and surroundings in real-time.

The zero-touch control plane for mmWave multi-hop routing leverages vehicle and environment information in the DT cyber world to perform several key tasks: extracting the connection topology, estimating the mmWave channel model for connectivity assessment, and utilizing graph analysis and prediction to select the most efficient communication route for current and future data transmission. A detailed explanation of the multi-hop routing algorithm used in the control plane can be found in Section \ref{section:DT-multihop-algo}.

By integrating real-time information exchange within the DT with the automated control plane for mmWave multi-hop routing calculation, the proposed system can ensure reliable and dynamic routing within the mmWave network.

\subsection{mmWave V2X channel model}
To evaluate the channel status of mmWave V2X in dynamic traffic environments, it is necessary to utilize a path loss/signal attenuation model that considers the influence of blockages:
\begin{equation}
PL(d)[dB]=10\rho\times \mathrm{log}_{10}(d)+\gamma+15d/1000,
\end{equation}
where $d$ is the length of a mmWave V2X link, $\rho$ is the mmWave attenuation coefficient, and $\gamma$ is a constant. The $15d/1000$ represents the 15 dB attenuation per kilometer of mmWave (60 GHz) in the atmosphere. The values of  $\rho$ and $\gamma$ are determined based on different mmWave V2X scenarios. The reference values are provided in \cite{cite:mmwave-model} for line-of-sight (LOS) V2X and non-light-of-sight (NLOS) V2X scenarios with various numbers of blockages. Using the environmental information in the DT, the path loss/signal attenuation can be estimated for any existing or potential mmWave V2X link.

\subsection{Reliability metric}
To evaluate the system's performance, we define the reliability metric as follows:
\begin{equation}
Reliability= \frac{\sum_{t=1}^{T} CV_{\text{connected}, t}}{\sum_{t=1}^{T} CV_{\text{total}, t}}, \label{eq:reliability-metric}
\end{equation}
where $CV_{\text{connected}, t}$ represents the number of connected vehicles that successfully satisfy their V2X connection demand at each timestep $t$. These vehicles can establish either direct communication or multi-hop communication to satisfy their demand. $CV_{\text{total}, t}$  denotes the total number of connected vehicles at timestep $t$, regardless of their connection status. This metric provides a quantitative measure of the system's reliability by proportionating the number of connected vehicles that successfully establish V2X communication over the total number of connected vehicles at each timestep.

\subsection{DT-based multi-hop routing algorithm \label{section:DT-multihop-algo}}

\begin{figure}[t]
    \centering

    \begin{subfigure}{0.482\linewidth}
        \includegraphics[width=\linewidth]{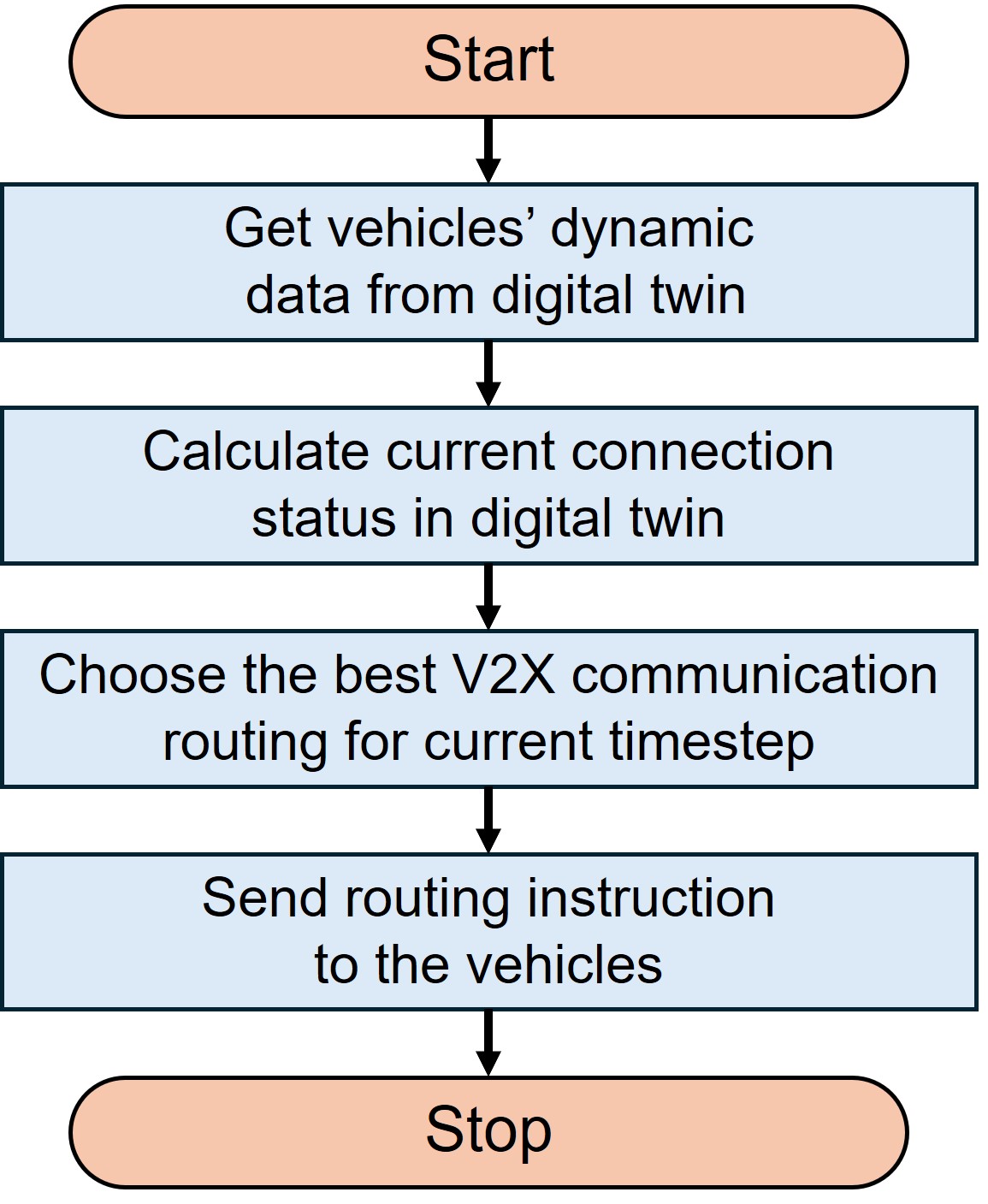}
        \caption{Real-time Routing \\ Calculation}
        \label{fig:real-time-analysis-flow}
    \end{subfigure}
    \hfill
    \begin{subfigure}{0.48\linewidth}
        \includegraphics[width=\linewidth]{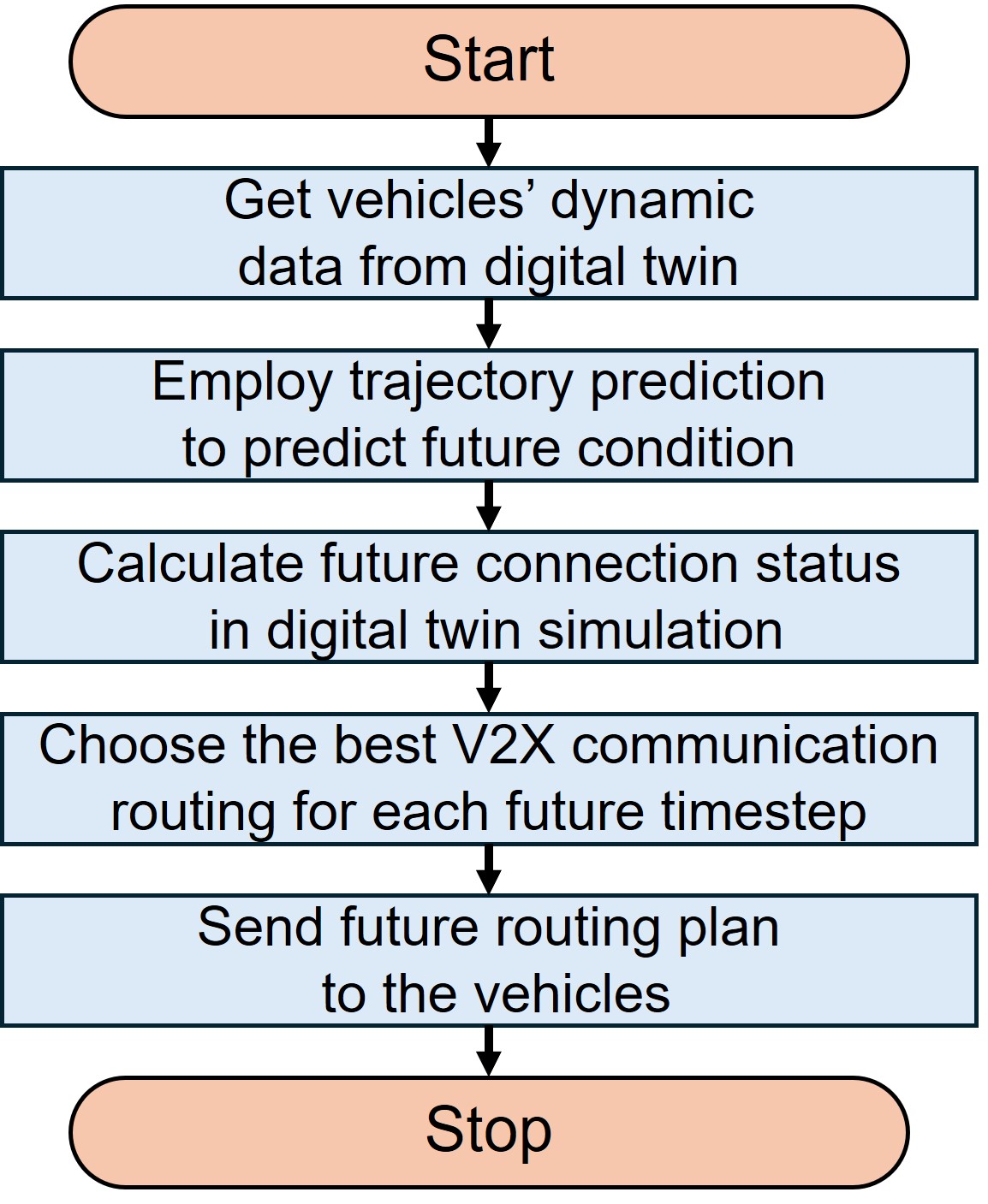}
        \caption{Future Route Planning \\ with Prediction}
        \label{fig:prediction-flow}
    \end{subfigure}
    
    \caption{
    DT-based multi-hop routing algorithm flowchart.}
    \label{fig:algorithms-flow}
\end{figure}

In this paper, we introduce two distinct multi-hop routing algorithms for the proposed system: real-time routing calculation and future route planning using trajectory prediction. Each algorithm flowchart is depicted in Fig \ref{fig:algorithms-flow}.

\paragraph{Real-time routing calculation}
In this strategy, the DT continuously processes real-time information in order to calculate optimal multi-hop routes for the current timestamp. As depicted in Fig. \ref{fig:real-time-analysis-flow}, at the initiation of each routing calculation cycle, the algorithm retrieves vehicles' dynamic data from the DT, including vehicle positions, speeds, traffic conditions, past connection status, and current connection demands. Based on the current state of the vehicles and the environment, the DT processes all possible current connection statuses and then chooses the best communication routing based on the calculated communication map. Once the optimal routes are determined, the DT issues communication instructions to the connected vehicles, instructing them on the most efficient mmWave communication routes for data transmission in real-time.

This real-time routing calculation approach offers the advantage of data freshness, allowing for the calculation of precise routing topology within each timestamp. However, the effectiveness of this real-time routing calculation strategy relies heavily on the end-to-end latency. This is primarily due to the dynamic nature of V2X communication, where conditions change rapidly in real time. When latency is high, the time taken for the DT to receive, process, and transmit data to vehicles increases, resulting in a delay between data acquisition and route calculation. Consequently, the calculated optimal route may be outdated when it reaches the vehicles, resulting in sub-optimal communication for the current conditions. 

\paragraph{Future route planning using trajectory prediction}
In this strategy, the DT leverages predictive analytics to forecast future traffic patterns and network conditions, facilitating proactive multi-hop route planning for future times. The algorithm flow is illustrated in Fig. \ref{fig:prediction-flow}. Firstly, this future route planning using trajectory prediction algorithm is employed at regular intervals. At the start of each run, the system retrieves current and historical data of vehicle states from the DT. Then, it uses a trajectory prediction algorithm to forecast future scenarios. By simulating connection statuses in potential future scenarios, the digital twin can identify optimal multi-hop routes for upcoming communication demands. Lastly, the digital twin proactively instructs vehicles on pre-planned routes well in advance. This enables vehicles to pre-establish communication links and optimize their routing strategies before actual demand arises.

This future route planning using trajectory prediction approach offers a significant advantage compared to previous real-time routing calculation methods. Instead of waiting for real-time updates, it sends planned communication routes to vehicles in advance, eliminating the necessity of depending on real-time latency. This approach enables vehicles to establish communication links and optimize their routing strategies beforehand, without the need for immediate instructions from the DT.

However, the reliability of this future route planning using trajectory prediction method depends on the accuracy of the trajectory prediction model. Since the future traffic scenarios in DT determine the reliability of the planned routes if the trajectory prediction models accurately forecast the future vehicle positions, the planned routes can closely align with ground truth future scenarios, resulting in efficient and reliable communication. Conversely, inaccurate prediction may lead to sub-optimal route planning and communication inefficiencies. 

\section{System Evaluation\label{section:evaluation}}
\subsection{Scenario \label{section:scenario}}
To validate the effectiveness of the proposed system architecture, this research conducts a simulation experiment considering line-of-sight (LOS) mmWave communication in an urban intersection scenario.

\begin{figure}[tb]
    \includegraphics[width=\linewidth]{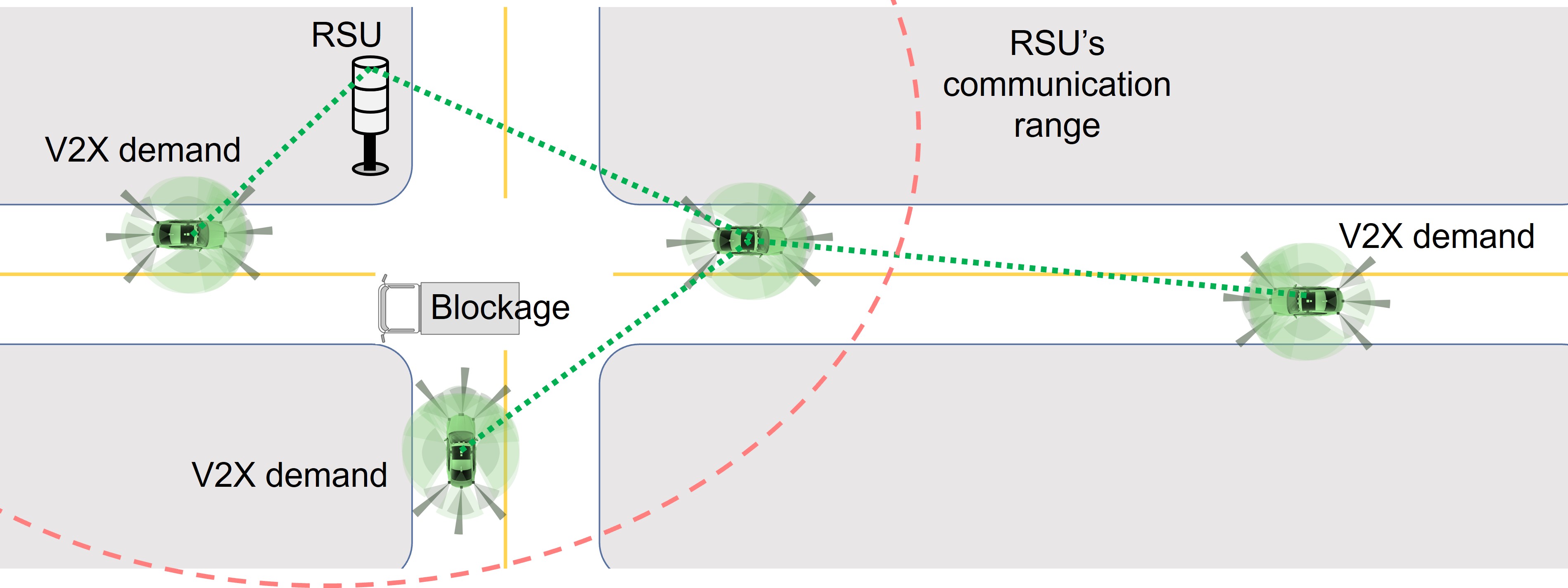}
    \centering
    \caption{System evaluation scenario.}
    \label{fig:scenario}
\end{figure}

In the system evaluation scenario depicted in Fig. \ref{fig:scenario}, we consider a single intersection in which a RSU is positioned at a height of 5 meters in the center of the intersection \cite{cite:rsu-height}. The scenario involves a combination of connected and unconnected vehicles navigating the area, with the objective for all connected vehicles to establish V2X communication with the RSU. This can be achieved either through a direct single hop to the RSU or by employing multi-hop communication techniques in cases where obstacles block direct communication or when vehicles are beyond the RSU's single-hop communication range.

This research conducted proof-of-concept experiments focusing on two distinct scenarios as follows:
\paragraph{Fully connected traffic scenario} All vehicles within the simulation have connectivity capabilities, enabling the ability to evaluate system connectivity under ideal conditions. 
\paragraph{Mixed traffic scenario} Presents both connected and unconnected vehicles, allowing for system performance measurement in the presence of blockages, as unconnected vehicles can act as moving obstacles obstructing LOS connections. \\
These distinct scenarios offer valuable insights into the system's performance under different conditions, providing a comprehensive evaluation of the proposed system.

\subsection{Simulation Implementation \label{section:sim-implement}}

\begin{figure}[tb]
    \includegraphics[width=\linewidth]{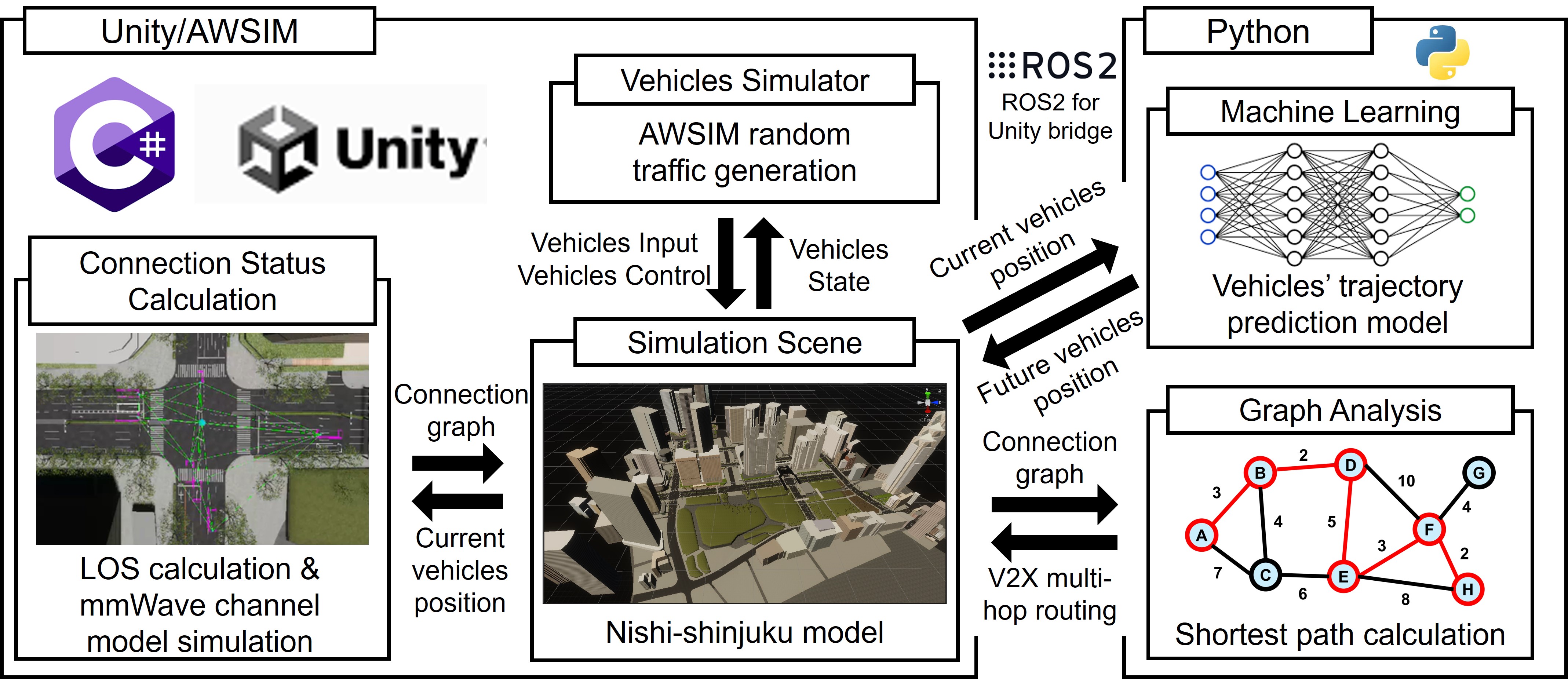}
    \centering
    \caption{Implementation of DT-based multi-hop routing management simulation system.}
    \label{fig:eval-sys-implement}
\end{figure}

The proposed DT-based multi-hop routing management system is evaluated using a simulation framework, as depicted in Fig. \ref{fig:eval-sys-implement}. This framework is composed of a traffic simulator and a physics simulator implemented in Unity, along with a graph processing and machine learning module implemented in Python.

Unity serves a dual purpose as both a traffic simulator for vehicle physical space and a digital twin for cyberspace itself. Utilizing the AWSIM package \cite{cite:awsim}, an open-source simulator designed for autonomous driving research and development, we simulated real-world traffic for the intersection in the Nishishinjuku area. Additionally, we incorporated Unity's ray-casting capabilities to simulate mmWave communication dynamics.

To implement the DT-based multi-hop management algorithms outlined in section \ref{section:DT-multihop-algo}, we leverage the Robot Operating System2 (ROS2) framework \cite{cite:ros2} to extend the calculation to Python. Within Python, we utilize the Dijkstra shortest path algorithm \cite{cite:dijkstra} to determine the optimal communication route and the multi-layer Long Short-Term Memory (LSTM) model \cite{cite:lstm} for vehicles' future trajectory prediction. By adopting this approach, we can effectively implement sophisticated routing algorithms beyond the limitations of Unity's native environment.



For reference, we also implement a conventional routing algorithm for comparison. This conventional method involves interval-based routing updates using the VANETs protocol. Taking into account the transmission delay between vehicles within VANETs, and decentralizing the calculation of V2X communication routing for each vehicle, we set the communication topology update interval to 5 seconds to accommodate this delay.


\subsection{Results}
\begin{figure}[t]
    \centering

    \begin{subfigure}{0.48\textwidth}
        \includegraphics[width=\linewidth]{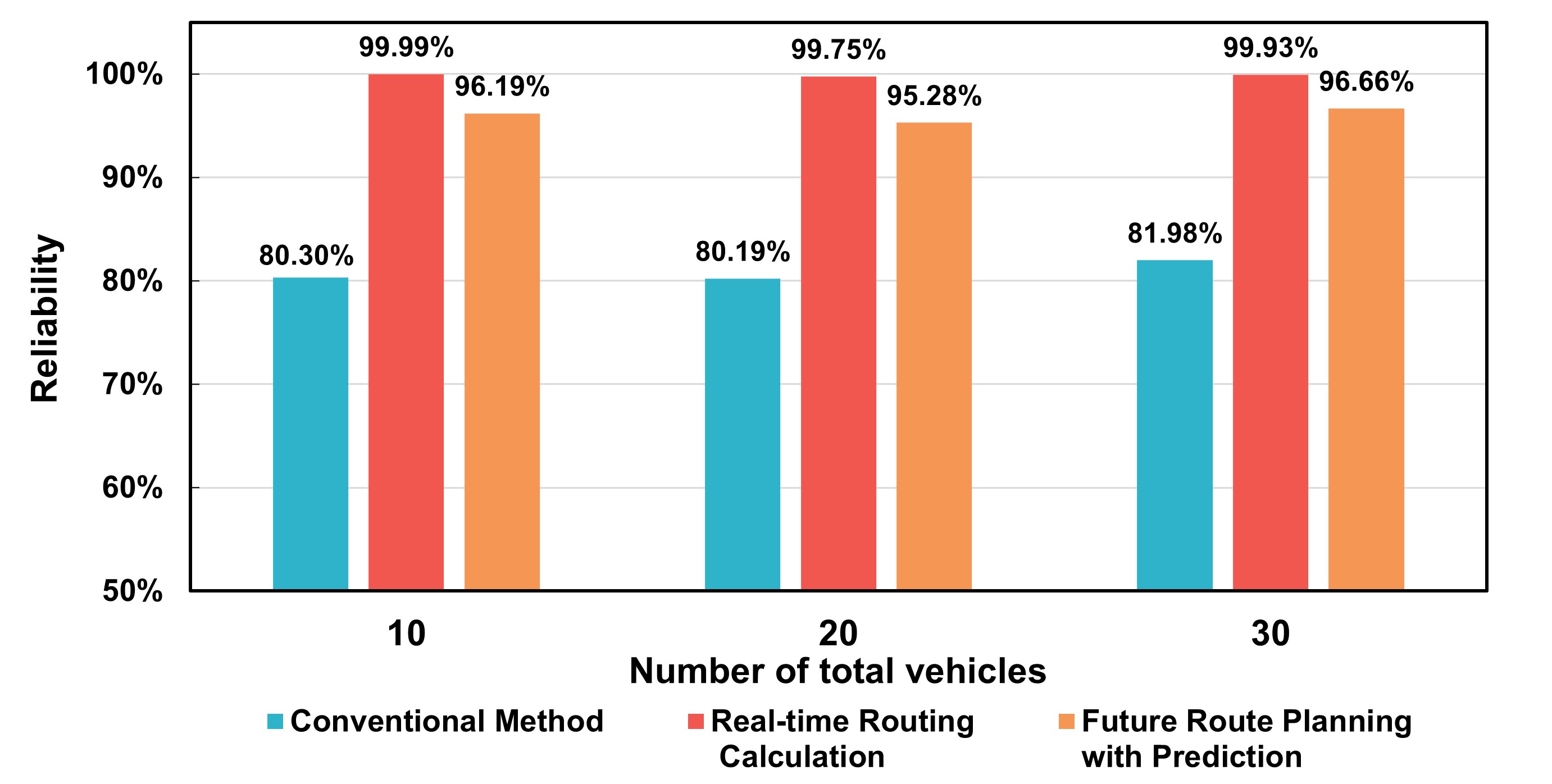}
        \caption{Fully connected traffic scenario}
        \label{fig:all-connected-res}
    \end{subfigure}
    \hfill
    \begin{subfigure}{0.48\textwidth}
        \includegraphics[width=\linewidth]{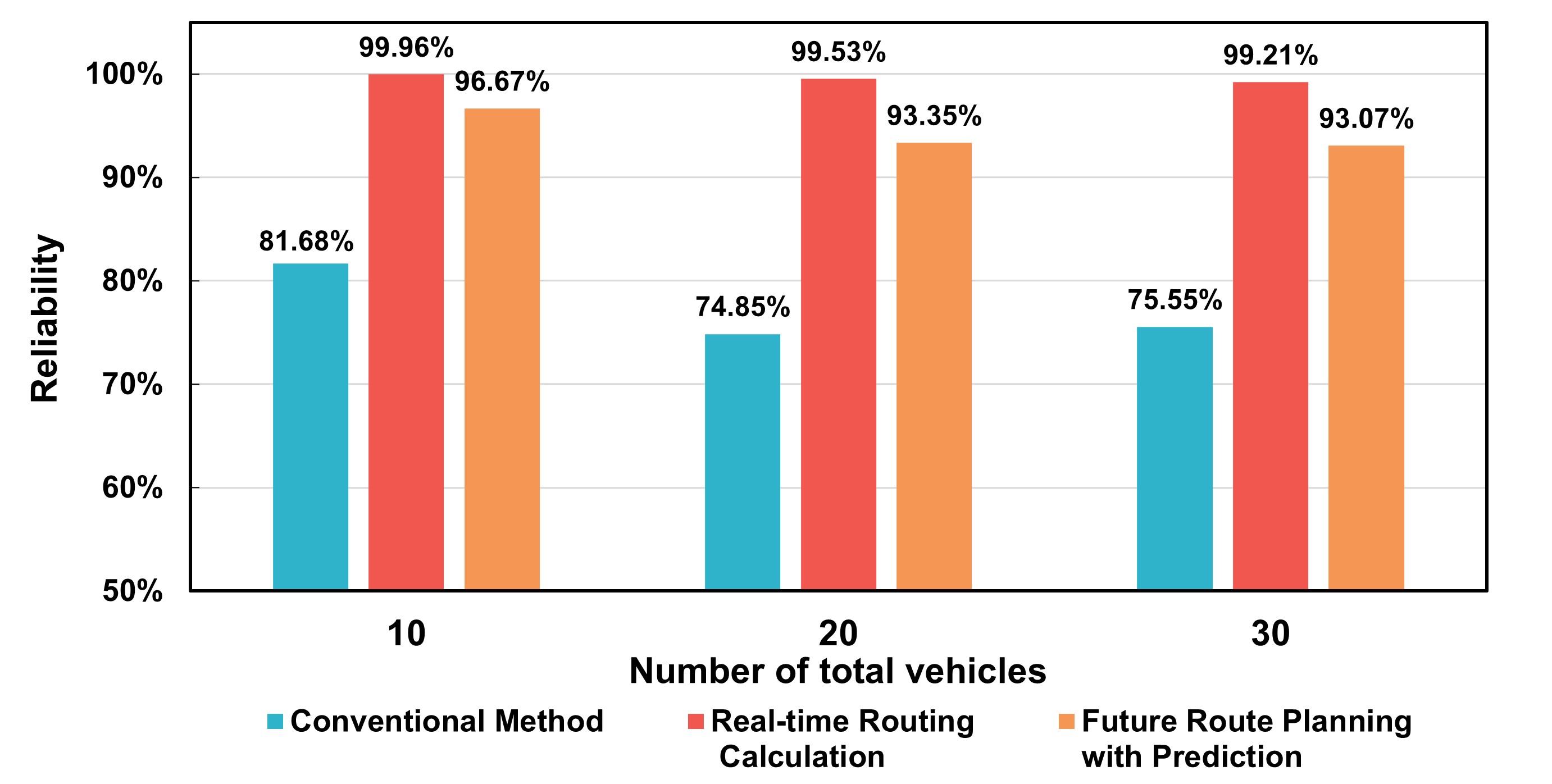}
        \caption{Mixed traffic scenario}
        \label{fig:mix-connected-res}
    \end{subfigure}

    \caption{System evaluation results.}
    \label{fig:sys-eval-res}
\end{figure}

Leveraging the evaluation system detailed in Section \ref{section:sim-implement}, we conducted a comprehensive reliability assessment for both the conventional and proposed routing methods across all scenarios outlined in Section \ref{section:scenario}. The results are presented in Fig. \ref{fig:sys-eval-res}. 

Fig. \ref{fig:all-connected-res} explores the relationship between communication reliability and the number of vehicles in the scenario where all vehicles are connected vehicles. As expected, with a low number of vehicles (10), communication is highly reliable (up to 99.99\%) due to minimal communication demand and minimal potential for blockage. However, when the number of vehicles increases to 20, introducing more potential blockage between vehicles, the maximum reliability drops slightly to 99.75\%. Interestingly, further increasing the number of connected vehicles to 30 improves reliability to 99.93\%. This is because the denser traffic of connected vehicles enhances communication coverage and enables multi-hop routing to overcome blockages.

The proposed DT-based multi-hop routing management method, particularly the real-time routing calculation algorithm, demonstrates clear dominance in reliability across all cases. In the 30-vehicle scenario, the conventional method achieves only 81.98\% reliability. In contrast, the proposed method's real-time routing calculation algorithm achieves near-perfect reliability (99.93\%), while the future route planning using trajectory prediction approach reaches 96.66\%.

Fig. \ref{fig:mix-connected-res} examines the impact of the mixed traffic scenario, where both connected and unconnected vehicles are present. The dynamic movement of unconnected vehicles introduces significant blockage, evident in the lower overall reliability compared to the all-connected case (Fig. \ref{fig:all-connected-res}). The maximum reliability reaches only 99.96\% and further declines as the number of vehicles increases. This is due to increased communication demand from connected vehicles and blockages caused by unconnected vehicles.

Similar to the all-connected scenario, the proposed method outperforms the conventional method in terms of reliability (Fig. \ref{fig:mix-connected-res}). Once again, the real-time routing calculation approach achieves the best results. For example, with 30 vehicles, it delivers a reliability of 99.21\%. The future route planning using trajectory prediction approach follows at 93.07\%, and the conventional method falls behind at only 75.55\% reliability.

In summary, the communication routing reliability evaluation (Fig. \ref{fig:sys-eval-res}) confirms the significant reliability advantage of the proposed DT-based routing method. This is due to the dynamic route updates of the proposed system, which overcome the conventional method's limitations of outdated topologies. The real-time routing calculation algorithm achieves exceptional reliability, reaching 99.93\% in the fully connected traffic scenario and 99.21\% even in environments with blockage. However, it is noteworthy that the simulation did not consider latency between the DT and real vehicles, which could potentially affect reliability in real-world scenarios. The future route planning using trajectory prediction approach emerges as the second-best performer in this evaluation. It achieves a reliability of 96.66\% in the fully connected scenario and 93.07\% in blocking situations. While its accuracy falls short of the real-time calculation approach, reflecting accuracy limitations in the ML-based vehicle trajectory prediction. However, unlike the real-time method, this future planning approach offers robustness to system end-to-end latency.

\section{Conclusion and Future Works\label{section:conclusion}}
In conclusion, this paper has utilized DT technology to enhance mmWave V2X reliability by introducing a zero-touch control plane for multi-hop routing management. The research presents two DT-based multi-hop routing algorithms employed by the control plane: real-time routing calculation for accurate routing and future route planning using trajectory prediction to mitigate latency issues. The Unity-based proof of concept in the urban intersection scenario showcased the effectiveness of the proposed system compared to traditional VANETs protocols in both fully connected and mixed traffic scenarios. The results demonstrate significant improvements in mmWave V2X reliability, offering practical solutions for future communication systems in dynamic traffic environments. Future work could explore reducing vehicle-DT communication latency, developing more accurate prediction algorithms, or implementing more sophisticated multi-hop routing algorithms for further optimization.

\section*{Acknowledgment}
This work was partly supported by DENSO Corporation.

\bibliographystyle{bib/IEEEtran}
\bibliography{bib/bibliography}

\end{document}